\documentclass[letterpaper]{article} 

\ifdefined\aaaianonymous
    \usepackage[submission]{aaai2026}  
\else
    \usepackage{aaai2026}              
\fi

\usepackage{times}  
\usepackage{helvet}  
\usepackage{courier}  
\usepackage[hyphens]{url}  
\usepackage{graphicx} 
\usepackage[numbers]{natbib}
\usepackage{caption} 
\usepackage{amsmath} 
\usepackage{algorithm}
\usepackage{algorithmic}
\usepackage{booktabs}
\usepackage{multirow}
\usepackage{siunitx}
\usepackage[table]{xcolor}
\usepackage{amsmath}
\usepackage{tabularx}
\usepackage{booktabs}
\usepackage{colortbl}

\usepackage{newfloat}
\usepackage{listings}
\usepackage{amssymb}
\usepackage{pifont}
\usepackage{booktabs}   
\usepackage{siunitx}    
\usepackage{xcolor}     
\usepackage{amsmath}    
\usepackage[svgnames]{xcolor}
\usepackage[framemethod=default]{mdframed}
\newmdenv[
    linecolor=gray!80!black,
    backgroundcolor=gray!10!white,
    frametitlebackgroundcolor=gray!80!black,
    frametitlerule=true,
    frametitlefont=\color{white}\bfseries,
    frametitle={Template for HiMed-RL},
    linewidth=0.5mm,
    innertopmargin=\topskip,
]{mybox}

\DeclareCaptionStyle{ruled}{labelfont=normalfont,labelsep=colon,strut=off} 
\floatstyle{ruled}
\newfloat{listing}{tb}{lst}{}
\floatname{listing}{Listing}

\ifdefined\aaaianonymous
    \title{Beyond N-grams: A Hierarchical Reward Learning Framework \\ for  Clinically-Aware Medical Report Generation}
\else
    \title{Beyond N-grams: A Hierarchical Reward Learning Framework \\ for  Clinically-Aware Medical Report Generation}
\fi

\author{
    Yuan Wang\textsuperscript{\rm 1},
    Shujian Gao\textsuperscript{\rm 2},
    Jiaxiang Liu\textsuperscript{\rm 1,4},
    Songtao Jiang\textsuperscript{\rm 1},
    Haoxiang Xia \textsuperscript{\rm 1},
    Xiaotian Zhang  \textsuperscript{\rm 1},
    Zhaolu Kang  \textsuperscript{\rm 3},
    Yemin Wang \textsuperscript{\rm 1},
    Zuozhu Liu \textsuperscript{\rm 1*}
}
\affiliations{
    \textsuperscript{\rm 1}Zhejiang University \quad
    \textsuperscript{\rm 2}Fudan University \quad
    \textsuperscript{\rm 3}Peking University\\
     \textsuperscript{\rm 4}Guangdong Institute of Intelligence Science and Technology\\
    
}

\usepackage{bibentry}   

\begin{document}

\maketitle

\begin{abstract}
Automatic medical report generation can greatly reduce the workload of doctors, but it is often unreliable for real-world deployment. Current methods can write formally fluent sentences but may be factually flawed, introducing serious medical errors known as clinical hallucinations, which make them untrustworthy for diagnosis. To bridge this gap, we introduce \textbf{HiMed-RL}, a Hierarchical Medical Reward Learning Framework designed to explicitly prioritize clinical quality. HiMed-RL moves beyond simple text matching by deconstructing reward learning into three synergistic levels: it first ensures linguistic fluency at the token-level, then enforces factual grounding at the concept-level by aligning key medical terms with expert knowledge, and finally assesses high-level diagnostic consistency at the semantic-level using a specialized LLM verifier. This hierarchical reward is implemented via a \textbf{Human-inspired Dynamic Reward Adjustment}, a strategy which first teaches the model to learn basic facts before progressing to more complex diagnostic reasoning. Experimentally, \textbf{HiMed-3B} achieves state-of-the-art performance on both in-domain and out-of-domain benchmarks, particularly on the latter, with an improvement of \textbf{12.1\%} over the second-best baseline. Our work provides a robust paradigm for generating reports that not only improve fluency but clinical fine-grained quality.

%



\end{abstract}

\ifdefined\aaaianonymous
\else
\begin{links}
    \link{Code}{https://github.com/Venn2336/HiMed-RL}
\end{links}
\fi

\section{Introduction}
\label{intro}
Automatic Medical Report Generation (MRG), which aims to generate textual descriptions from medical images, is a promising solution to alleviate the documentation burden in clinical practice~\citep{medregion,guo2024automatic}. More than just writing fluently, reports must be factually accurate with the visual information and show a deep understanding of medical knowledge~\cite{seedetail}. However, current methods often struggle to meet these clinical standards.

Conventional MRG-specific methods, such as R2Gen~\cite{r2gen} and Att2in~\cite{xu2016showattendtellneural}, tend to overfit the training data; while achieving high keyword overlap, they often underperform on clinical metrics that assess semantic accuracy and medical relevance~\cite{li2025joint}. 
Besides, another line of research has focused on refining the language modeling process within the Supervised Fine-Tuning (SFT) paradigm for Multi-modal Large Language Models (MLLMs)~\cite{liu2024medcot,jiang2025hulu}. 
These efforts include advanced prompt engineering with medical entities~\cite{promptmrg} or question-driven cues~\cite{mepnet}, and enhanced fine-tuning strategies to improve visual comprehension~\cite{seedetail}.
However, such methods are fundamentally bottlenecked by the SFT objective itself. The standard goal of maximizing token-level likelihood~\citep{zeng2024token} does not inherently enforce the factual integrity and logical consistency that are paramount for deployment in clinical practice.

\begin{figure}[ht]
    \centering
    \includegraphics[width=0.96\linewidth]{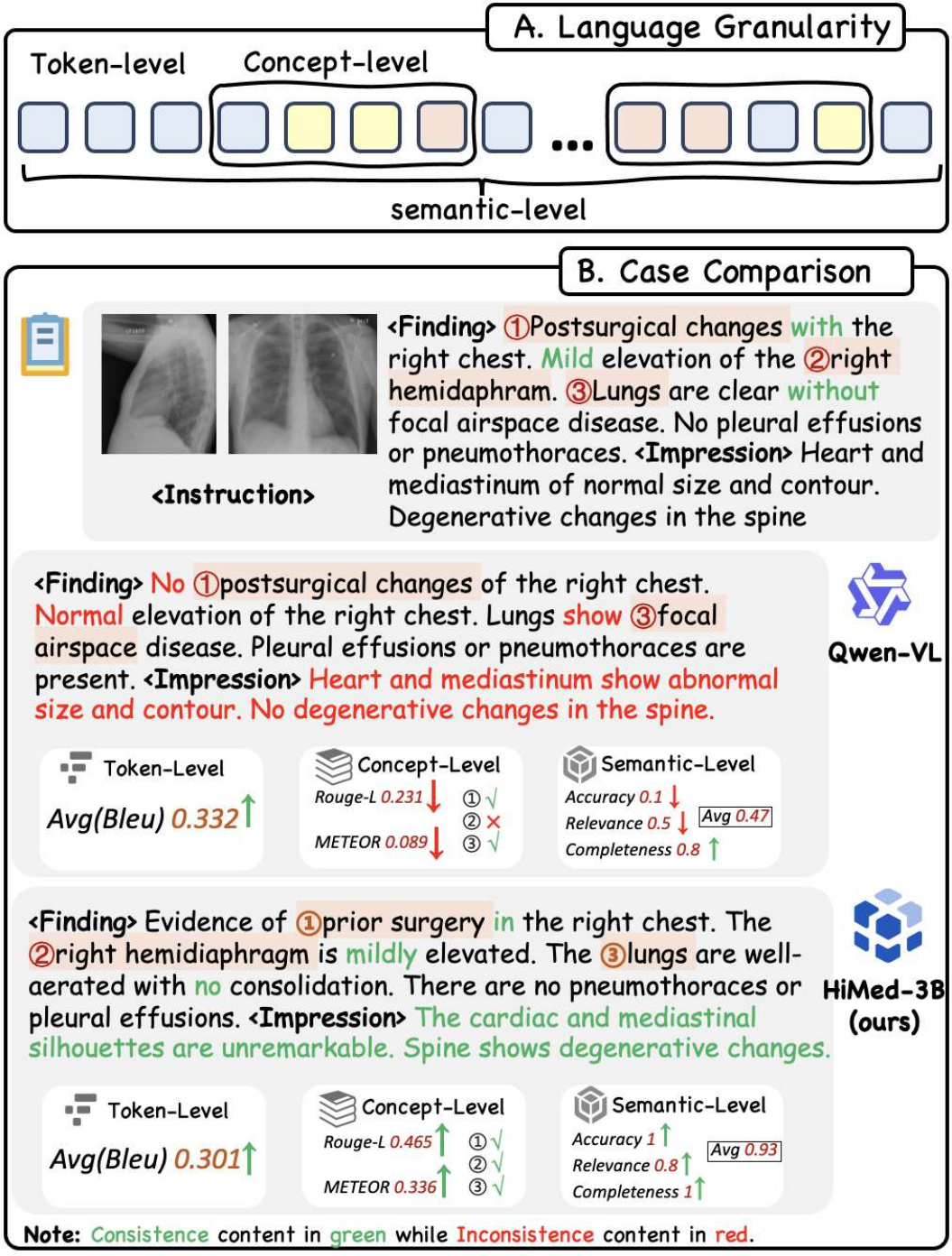}
    \caption{(a) \textit{Token-level}, the report text is decomposed into individual words or sub-word tokens for assessment. \textit{Concept-level}, relevant tokens are aggregated into clinically significant medical concepts, e.g., postsurgical changes. \textit{Semantic-level} aims to comprehend the core meaning and diagnostic conclusion of the entire report. (b) Our HiMed-3B model outperforms Qwen2.5-VL-7B across three evaluation levels: at the token-level, with a comparable average BLEU score; at the concept-level, by correctly identifying three medical entity types and achieving superior ROUGE-L and METEOR scores; and at the semantic-level, demonstrating enhanced accuracy, relevance, and completeness.
     }
    \label{fig:enter-label}
\end{figure}



Reinforcement Learning (RL) has demonstrated exceptional capabilities in generalization, complex reasoning, and trustworthy alignment, making it a highly suitable paradigm for MRG. However, the potential of RL in this domain is currently hindered by simplistic reward designs. Current methods rely on rule-based rewards that operate at lower linguistic levels, mainly focusing on token and concept matching. As we illustrate in \textbf{\textit{Figure~\ref{fig:enter-label}(a)}}, linguistic quality in MRG can be assessed at three levels: token, concept, and semantic. For instance, token-level approaches often depend on superficial metrics like BLEU~\citep{zou2025corbenchxlargescalechestxray}. At the concept-level, a significant body of work incentivizes matching medical keywords to improve clinical utility~\citep{zhang2025medtvtr1multimodalllmempowering, dai2025qoqmedbuildingmultimodalclinical,fan2025chestxreasoneradvancingradiologyfoundation,zhang2025med}.

However, we argue that the complexity of MRG necessitates a semantically coherent and procedurally trustworthy reasoning process, which cannot be adequately addressed by the aforementioned reward types alone. 
Previous reward designs fall short, particularly in generating clinically consistent reports \cite{icon}, mitigating factual hallucinations \cite{medkp,xu2023hybrid}, resolving semantic contradictions \cite{zhou2021visual}, and perceiving underlying pathological structures \cite{dynamic}, since rule-based mechanisms struggle to evaluate these challenges directly.

Hence, we introduce \textbf{HiMed-RL}, a hierarchical reward learning paradigm for MRG. 
Specifically, as shown in \textbf{\textit{Figure~\ref{fig:enter-label}}}, at the token-level, we ensure linguistic fluency and syntactic correctness to establish report readability. 
Next, at the concept-level, our reward encourages alignment with key medical terminology and domain knowledge, reinforcing entity consistency and factual grounding to mitigate clinical hallucinations. 
In contrast to previous works \cite{fan2025chestxreasoneradvancingradiologyfoundation}, our method introduces an LLM verifier to conduct a high-level evaluation of the semantic integrity and diagnostic consistency of the entire report, ensuring the rationality of the reasoning logic and the credibility of the medical judgment.
In terms of training strategy, inspired by the real-scenario radiologist-decision process, we use dynamic reward weight adjustment, starting with learning basic facts and gradually transitioning to training more challenging diagnostic reasoning abilities. 
This phased optimization method not only improves learning stability but also more closely follows the physician's thought process of \textit{from factual statement to clinical judgment} when writing reports. Our contributions are three-fold:

\begin{itemize}
    \item We reconceptualize MRG as a complex long sequence modeling task. Unlike the previous textual-similarity–driven optimization paradigm, we introduce \textbf{HiMed-RL}, a framework that jointly optimizes reward signals at the token, concept, and semantic levels.
    \item We introduce \textbf{Human-inspired Dynamic Reward Adjustment} policy  during training. This mechanism guides the model through a progressive optimization process, transitioning from foundational linguistic expression to complex medical reasoning, thereby mirroring the cognitive workflow of clinicians in real-world scenarios.

    \item We conduct extensive experiments to demonstrate that our proposed \textbf{HiMed-3B} achieves state-of-the-art (SOTA) performance on multimodal medical imaging datasets. Our detailed evaluations validate the effectiveness and superiority of our approach.
\end{itemize}




\section{Related Work}
\subsection{Medical Report Generation}



MRG, a specialized sub-task of vision-language learning, aims to automatically generate coherent and accurate diagnostic reports from medical images.
Prior efforts have sought to enhance model performance by integrating external structures and knowledge. These strategies include embedding cognitive modules into the network architecture~\cite{archiecture,huang2025need}, leveraging medical knowledge graphs~\cite{kargen}, and employing multi-task learning to foster more robust and clinically-aware representations~\cite{promptmrg}.
However, those methods are inherently limited by their reliance on training from scratch with small-scale medical datasets, which restricts generalization. The advent of LLMs marks a paradigm shift~\cite{ma2025towards}; the extensive pre-training on vast corpora provides robust reasoning and fluency that can overcome the data scarcity issues to the medical domain, fundamentally reshaping the technical landscape for MRG.

\subsection{Reward Learning}
Reward learning has significantly advanced the reasoning and generation capabilities of LLMs, enhancing the out-of-domain generalization. 
Nevertheless, designing an appropriate reward learning framework tailored to the unique demands of different tasks remains a challenge. This often requires task-specific reward engineering.
For instance, text summarization may employ rewards for information overlap alongside penalties for excessive length~\citep{aggarwal2025l1controllinglongreasoning}; Question-Answering tasks frequently rely on accuracy-based rewards~\citep{wang2025v2t}. In this work, we address this challenge within the context of MRG.

\section{Methodology}

\begin{figure*}[h]
    \centering
    \includegraphics[width=1\linewidth]{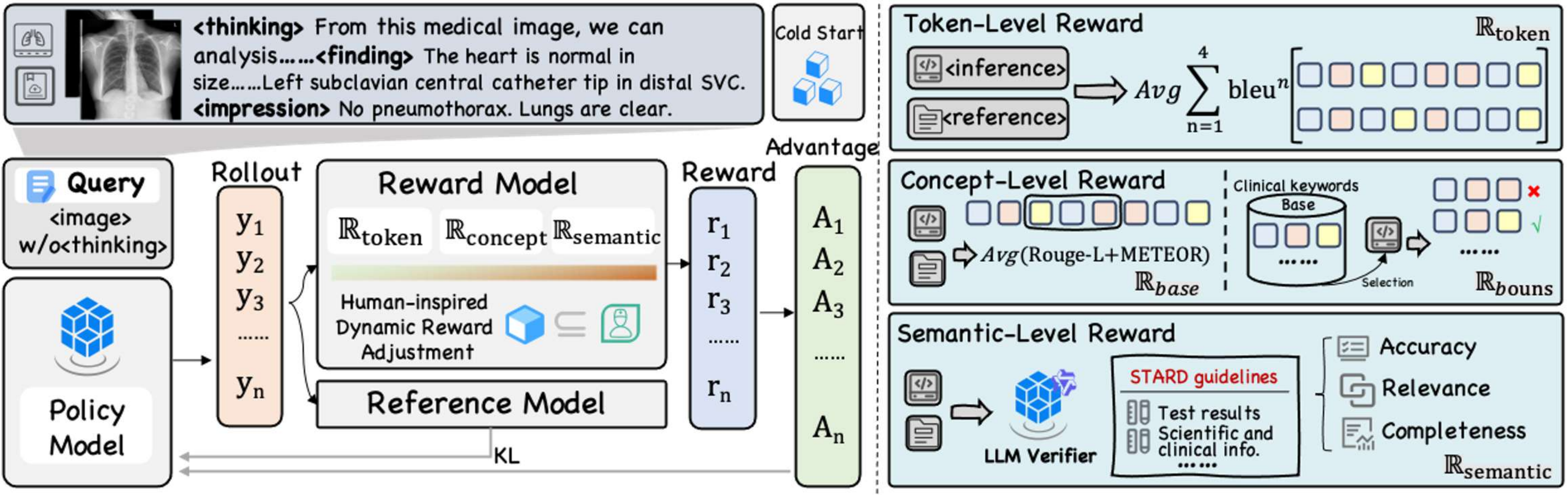}
    \caption{An overview of the HiMed-RL pipeline. MRG task is trained by a Hierarchical Reward Learning Framework that integrates token-level, concept-level, and semantic-level rewards. Human-inspired Dynamic Reward Adjustment strategy guides the model's learning process, transitioning from foundational fluency to complex medical reasoning, to optimize report quality.}
    \label{fig:pipeline}
\end{figure*}

In this section, we present our approach, which employs RL with a Hierarchical Medical Reward Learning Framework and Human-inspired Dynamic Reward Adjustment policy, as shown in \textbf{\textit{Figure \ref{fig:pipeline}}}.

\subsection{Hierarchical Medical Reward Scaling}
\label{sec:method_reward}

As discussed in the introduction, current reward designs for MRG cannot sufficiently address issues such as clinical hallucinations, semantic contradictions, and overall coherence. This is primarily because a comprehensive reward learning framework for MRG is still absent. To address this limitation, we propose Hierarchical Reward Learning strategy. (Details of prompt design are given in \textbf{\textit{Appendix A.2}})

\subsubsection{Preliminaries and Formal Definitions}

In our setting, given a medical image~$I$, the goal of our policy model~$\pi$~is to generate a piece of text description, that is, a medical report~$Y$. We formalize both the generated report~$Y$~and the gold standard reference report~$\hat{Y}$~as a sequence of tokens:
\begin{itemize}
\item \textbf{Generated Report:} $Y=(y_1, y_2, \dots, y_L)$, with a sequence length of $L$.
    \item \textbf{Reference Report:} $\hat{Y}=(\hat{y}_1, \hat{y}_2, \dots, \hat{y}_{\hat{L}})$, with a sequence length of $\hat{L}$.
\end{itemize}

For the precise definition of subsequent reward functions, we first introduce the concept of n-gram. An n-gram is a continuous subsequence of length n extracted from a token sequence. We use~$\text{grams}_n(S)$~to denote the set of all n-grams extracted from an arbitrary sequence~$S$.

\subsubsection{Token-Level Reward ($\mathbb{R}_\text{token}$)}

It is designed to measure the linguistic similarity between a generated sequence $Y$ and a reference sequence $\hat{Y}$. Its formulation is inspired by the BLEU score and integrates two key components: modified n-gram precision and a brevity penalty.


A simple calculation of n-gram precision measures the fraction of n-grams in the candidate sequence $Y$ that appear in the reference sequence $\hat{Y}$. However, this approach improperly handles repeated words, as a model could be rewarded for the over-generation of high-frequency terms. Therefore, we employ a modified n-gram precision, denoted as $p_n$, which incorporates a clipping mechanism. 
\begin{equation}
    p_n(Y, \hat{Y}) = \frac{\sum_{g \in \text{grams}_n(Y)} \min(C(g, Y), C(g, \hat{Y}))}{\sum_{g' \in \text{grams}_n(Y)} C(g', Y)},
\end{equation}
where $\text{grams}_n(S)$ represents the multiset of n-grams in a sequence $S$, and $C(g, S)$ denotes the count of a specific n-gram $g$ within $S$.

In addition, to penalize generated sequences that are unduly shorter than their references, we introduce a brevity penalty (BP). This factor is defined as:
\begin{equation}
    \text{BP} = \min\left(1, \exp\left(1 - \frac{\hat{L}}{L}\right)\right),
\end{equation}
where $L$ and $\hat{L}$ are the respective lengths of the generated sequence $Y$ and the reference sequence $\hat{Y}$.

The final reward function $\mathbb{R}_{\text{token}}$ combines these elements:
\begin{equation}
    \mathbb{R}_{\text{token}}(Y, \hat{Y}) = \text{BP} \cdot \sum_{n=1}^{4} \lambda_n p_n(Y, \hat{Y}),
\end{equation}
in our implementation, we use the standard BLEU-4 formulation, where $\lambda_n$ are weights for each n-gram order.

\subsubsection{Concept-Level Reward ($\mathbb{R}_\text{concept}$)}

Above the form, $\mathbb{R}_{\text{concept}}$ aims to assess the consistency of the generated sequence~$Y$~with~$\hat{Y}$~in clinical factual content. It is composed of a base alignment reward~$R_{\text{base}}$~and a constrained entity reward bonus~$R_{\text{bonus}}$. The former is measured through ROUGE-L and METEOR to evaluate the coverage and accuracy of basic clinical concepts,
\begin{equation}
    \mathbb{R}_{\text{base}}(Y, \hat{Y}) = F_{\text{LCS}}(Y, \hat{Y}) + \text{METEOR}(Y, \hat{Y}),
\end{equation}
where $F_{\text{LCS}}$ is the F1-score component of ROUGE-L, calculated using the Longest Common Subsequence (LCS) between the candidate sequence $Y$ and reference $\hat{Y}$.

To encourage the generation of key medical entities, we introduce a bonus reward, $\mathbb{R}_{\text{bonus}}$. This reward is granted based on a predefined set of critical keywords, $\mathcal{K}$, which includes terms from USMLE~\footnote{United States Medical Licensing Examination} and RSNA~\cite{langlotz2006radlex} standardized medical vocabularies. A reward is given for each keyword $k \in \mathcal{K}$ (represented as a token sequence) found as a subsequence in the output $Y$. To prevent the model from exploiting this reward through simple repetition, the total bonus is capped at a maximum value of $\tau_{\text{limit}}$.


Finally, the total concept-level reward combines the base and bonus components as follows:
\begin{equation}
    \mathbb{R}_{\text{concept}}(Y, \hat{Y}) = \mathbb{R}_{\text{base}}(Y, \hat{Y}) + \min\left( \sum_{k \in \mathcal{K}} \beta \cdot \mathbb{I}(k \subseteq Y), \tau_{\text{limit}} \right),
\end{equation}
where $\mathbb{I}(k \subseteq Y)$ is an indicator function that returns 1 if keyword $k$ is found in the output $Y$, and $\beta$ is the reward value for a single match.

\subsubsection{Semantic-Level Reward ($\mathbb{R}_\text{semantic}$)}
While token-level and concept-level rewards ensure linguistic fluency and factual accuracy of individual clinical entities, they are insufficient for evaluating the overall clinical coherence and diagnostic integrity of the entire report. To bridge this gap, we introduce a semantic-level reward, $R_\text{semantic}$, which leverages a powerful clinical verifier as an impartial judge to perform a holistic evaluation of the generated report. This LLM verifier assesses the report's quality along three critical axes of clinical utility: Accuracy, Relevance, and Completeness, grounded in the Standards for Reporting of Diagnostic Accuracy (STARD) 2015 guidelines \cite{cohen2016stard}.


We formalize the scoring function as $\mathcal{J}$, which takes the generated report $Y$ and the ground-truth reference report $\hat{Y}$ as inputs. It then outputs a vector of scores, where each score is normalized to the range $[0, 1]$.
\begin{equation}
\mathcal{J}(Y, \hat{Y})
= s_{\text{acc}} + s_{\text{rel}} + s_{\text{com}},\quad
s_{\text{acc}},\, s_{\text{rel}},\, s_{\text{com}} \in [0,1].
\end{equation}

The evaluation dimensions are defined as follows:

\begin{itemize}
    \item \textbf{Accuracy ($s_{\text{acc}}$):} Assesses factual correctness by penalizing clinical \textit{hallucinations} (unfounded findings) and \textit{contradictions} against the reference report. This metric is guided by STARD's principles of reporting diagnostic accuracy (Items 1, 24) and systematically comparing results against a reference standard (Item 23).

    \item \textbf{Relevance ($s_{\text{rel}}$):} Measures clinical pertinence by evaluating whether the report prioritizes critical pathological findings and avoids verbose, non-contributory descriptions. This aligns with STARD's emphasis on defining the test's clinical role (Item 3) and its implications for practice (Item 27).

    \item \textbf{Completeness ($s_{\text{com}}$):} Quantifies the coverage of all essential clinical observations from the reference, penalizing the omission of significant findings. Inspired by STARD's requirement for comprehensive data reporting (Items 19-21, 23).
\end{itemize}

For semantic-level reward evaluation, our prompt template follow the STARD guidelines, which includes both the generated report ($Y$) and the reference report ($\hat{Y}$). (Details of LLM Verifier Case Study will be seen in \textbf{\textit{Appendix A.3}}).

\paragraph{Format Reward} We use regular expression extraction to enforce a structured response format. The model is required to place its reasoning process within \texttt{<think></think>} tags and provide the medical report inside \texttt{<finding></finding>} and \texttt{<impression></impression>} tags. The format reward score (\( \mathbb{R}_\text{format} \)) is computed as:
\begin{equation}
\mathbb{R}_\text{format} =
\begin{cases}
\text{1}, & \text{if format is correct} \\
\text{-1}, & \text{if format is incorrect}
\end{cases}
\end{equation}

\subsubsection{Total Reward Function}
To synergistically optimize for both foundational correctness and high-level clinical reasoning, we construct a composite total reward, $\mathbb{R}_{\text{total}}$. 
First, Low-Level Reward $\mathbb{R}_{\text{low-level}}$, which assesses structural and factual integrity, is formulated below:
\begin{equation}
    \mathbb{R}_{\text{low-level}} = w_t \mathbb{R}_{\text{token}} + w_c \mathbb{R}_{\text{concept}} + w_f \mathbb{R}_{\text{format}},
    \label{eq:low_level_reward}
\end{equation}
where $w_t, w_c, \text{and } w_f$ are scalar weights.

The total reward, $\mathbb{R}_{\text{total}}$, is then defined as a dynamically weighted combination of this low-level reward and the high-level semantic reward, $\mathbb{R}_{\text{semantic}}$. At any given training step $t$, the function is:
\begin{equation}
    \mathbb{R}_{\text{total}}(t) = \alpha_1(t) \cdot \mathbb{R}_{\text{low-level}} + \alpha_2(t) \cdot \mathbb{R}_{\text{semantic}}
    \label{eq:total_reward}
\end{equation}
The time-varying hyperparameters, $\alpha_1(t)$ and $\alpha_2(t)$, are crucial for our training strategy, enabling a strategic shift in optimization focus as training progresses.

\subsection{Human-inspired Dynamic Reward Adjustment}

A critical question arises within a multi-level reward learning framework: how can we orchestrate its various components to foster a coherent learning progression? Hence, we propose the Human-inspired Dynamic Reward Adjustment policy, which guides the model through a progressive learning process, transitioning from mastering fundamental concepts to performing complex reasoning. The adjustment mechanism is scheduled as follows:
\begin{itemize}
    \item Initial Phase: At the beginning of the training, we set a high value for $\alpha_1(t)$ and a low value for $\alpha_2(t)$. This encourages the model to focus on generating linguistically fluent and factually accurate text, mastering the building blocks of a valid report.
    \item Transition Phase: As training progresses, we gradually decrease $\alpha_1(t)$ while simultaneously increasing $\alpha_2(t)$ We employ a linear scheduling function to ensure a smooth transition:
\end{itemize}
\begin{equation}
    \alpha_1(t) = \max\left(1 - \frac{t}{T}, \alpha_{\text{min}}\right), \quad \alpha_2(t) = 1 - \alpha_1(t)
    \label{eq:reward_adjustment}
\end{equation}

\noindent \textbf{RL Algorithm} We employ the Group Reward Policy Optimization (GRPO) algorithm~\citep{shao2024deepseekmathpushinglimitsmathematical} for training. 
For each input $q$, we sample a group of candidate outputs $\{o_i\}_{i=1}^G$, compute their advantages $A_i$ based on our rule-metric mixed rewards, and then optimize the policy $\pi_{\theta}$ by maximizing the GRPO objective function:
\begin{equation}
\begin{aligned}
J_{\mathrm{GRPO}}(\theta) 
&= \mathbb{E}_{q \sim P(Q),\, \{o_i\}_{i=1}^G \sim \pi_{\theta_{\mathrm{old}}}(O \mid q)} \\
&\Biggl[
  \frac{1}{G} \sum_{i=1}^G
  \min\!\Bigl(
    \frac{\pi_{\theta}(o_i \mid q)}{\pi_{\theta_{\mathrm{old}}}(o_i \mid q)}\,A_i,\, \\
    &\mathrm{clip}\!\Bigl(
      \frac{\pi_{\theta}(o_i \mid q)}{\pi_{\theta_{\mathrm{old}}}(o_i \mid q)},
      1-\varepsilon,\,
      1+\varepsilon
    \Bigr)
    A_i
  \Bigr)  \\
  &-\,\beta\,D_{\mathrm{KL}}\bigl(\pi_{\theta}\,\big\|\,\pi_{\mathrm{ref}}\bigr)
\Biggr],
\end{aligned}
\label{eq1}
\end{equation}
where $\varepsilon$ is the clipping hyperparameter and $\beta$ controls the KL divergence penalty against a reference policy $\pi_{\mathrm{ref}}$. Our implementation and hyperparameter settings follow the original work~\citep{shao2024deepseekmathpushinglimitsmathematical}.

\section{Experiment}
In our experiments, we aim to answer three core questions:

\begin{enumerate}
    \item[\textbf{Q1}] Does our proposed model outperform contemporary baselines, including MRG-specific methods and MLLMs-based MRG methods? \textbf{\textit{A1: Main Results.}}

    \item[\textbf{Q2}] Does our hierarchical reward design, emphasizing multiple linguistic granularities, enhance the generation of coherent, consistent, and semantically aligned medical reports to the ground truth? \textbf{\textit{A2: Ablation Study.}}

    \item[\textbf{Q3}] Does our Human-inspired Dynamic Reward Adjustment strategy effectively align the generation process of HiMed-RL with the diagnostic workflow of human radiologists, from basic facts interpretation to comprehensive report synthesis? \textbf{\textit{A3: Discussion and Case Study.}}
\end{enumerate}

\subsection{Experiment Settings}


\noindent \textbf{Baselines and Datasets.} We benchmarked \mbox{HiMed-RL} against four categories of baselines: (1) general multimodal models such as Qwen2.5-VL~\cite{bai2025qwen2} and InternVL3~\cite{zhu2025internvl3}; (2) medically fine-tuned models including LLaVA-Med~\cite{li2023llava} and MedGemma~\cite{sellergren2025medgemma}; (3) reinforcement learning-based methods like QoQMed-VL~\cite{dai2025qoq} and MedVLM-R1~\cite{pan2025medvlm}; (4) MRG-specific methods like R2Gen~\cite{r2gen}, RGRG~\cite{Tanida_2023}. We used MIMIC-CXR~\cite{johnson2019mimic}, Chexpert~\cite{irvin2019chexpert}, and IU-Xray~\cite{demner2015preparing} for in-domain evaluation, and Padchest-GR~\cite{de2025padchest} for out-of-domain generalization test (see \textbf{\textit{Appendix A.1}} for details).

\noindent \textbf{Evaluation Metrics.} To ensure a thorough and multi-faceted evaluation, we adopt a comprehensive suite of metrics organized across three distinct levels of granularity for a complementary and detailed assessment of report quality. Specifically, at the token-level, we employ BLEU-1 through BLEU-4 to measure n-gram precision and fluency. For concept-level, METEOR can consider precise word matching, including synonyms and rewriting, and provide a more detailed report quality assessment~\cite{banerjee2005meteor}. At the semantic level, clinical accuracy was assessed using the RaTEScore metric, which is specifically designed for the evaluation of medical reports. This metric prioritizes the accuracy of medical entities, such as anatomical details and diagnostic findings~\cite{zhao2024ratescoremetricradiologyreport}.


\noindent \textbf{Implementation Details.} We implement our HiMed-3B model and its training pipeline using the \texttt{verl} framework. The backbone is initialized from Qwen2.5-VL-Instruct-3B, and then undergoes supervised fine-tuning on the target medical datasets as a cold-start phase. For the reinforcement learning stage, we adopt GRPO algorithm. The learning rate is set to $1 \times 10^{-6}$, with a mini-batch size of 32. At each GRPO step, we sample a group of $G=16$ candidate reports for each set of medical images. The PPO clipping threshold is set to $\varepsilon = 0.2$, and the KL-divergence penalty weight is set to $\beta = 0.1$. For the semantic-level reward, we employ \texttt{qwen3-30b-a3b} as the LLM verifier. All experiments are conducted on a cluster of eight NVIDIA A100 GPUs. 
Hyperparameters settings are given in \textbf{\textit{{Appendix A.8}}}.

\begin{table}[htbp]
    \centering
    \small 
    \setlength{\tabcolsep}{2pt} 
    \begin{tabular}{@{}ccc ccc@{}}
        \toprule
        \multicolumn{3}{c}{\textbf{Components}} & \multicolumn{3}{c}{\textbf{Performance Metrics}} \\ 
        \cmidrule(r){1-3} \cmidrule(l){4-6}
        $w/ \mathbb{R}_\text{semantic}$ & $w/ \mathbb{R}_\text{concept}$ & $w/ \mathbb{R}_\text{token}$  & MIMIC-CXR & CheXpert & IU-Xray \\
        \midrule
        & &  & 0.050 & 0.054 & 0.073 \\
        & & \checkmark   & 0.192 & 0.129 & 0.263 \\
        & \checkmark &   & 0.191 & 0.117 & 0.289 \\
        \checkmark &  &   & 0.101 & 0.125 & 0.288 \\
        & \checkmark  & \checkmark & 0.258 & 0.138 & 0.301 \\
        \checkmark  & \checkmark & \checkmark & \textbf{0.271} & \textbf{0.157} & \textbf{0.319} \\
        \bottomrule
    \end{tabular}
    \caption{Ablation study of different components performance. We use Rouge-L to evaluate each datasets.} 
    \label{tab:ablation}
\end{table}

\begin{table}[htbp]
    \centering
    \small 
    \setlength{\tabcolsep}{4pt} 
    \begin{tabular*}{\columnwidth}{@{\extracolsep{\fill}} l ccc} 
        \toprule
        \textbf{Configuration} & \textbf{ROUGE-L} & \textbf{METEOR} & \textbf{RATE} \\
        \midrule
        \textit{No Adjustment (Fixed)} \\ 
        \quad $\alpha_1=0.5, \alpha_2=0.5$ & 0.264 & 0.211 & 0.525 \\
        \midrule
        \textit{Linear Decay Adjustment} \\
        \quad $T=5k, \alpha_{\text{min}}=0.1$ & 0.266 & 0.214 & 0.536 \\
        \quad $T=20k, \alpha_{\text{min}}=0.1$ & 0.269 & 0.218 & 0.539 \\
        \quad $T=10k, \alpha_{\text{min}}=0.0$ & 0.268 & 0.216 & 0.541 \\
        \quad $T=10k, \alpha_{\text{min}}=0.3$ & \textbf{0.272} & 0.219 & 0.535 \\
        \hline
        \quad \textbf{$T=10k, \alpha_{\text{min}}=0.1$} & 0.271 & \textbf{0.220} & \textbf{0.544} \\
        \bottomrule
    \end{tabular*}
    \caption{Impact of Human-inspired Dynamic Reward Adjustment hyperparameters on the MIMIC-CXR. Our final configuration is highlighted.}
    \label{tab:reward_adjustment_compact}
\end{table}


\begin{table*}[ht]
\centering
\resizebox{0.95\linewidth}{!}{
\begin{tabular}{clccccccccccc}
\toprule
 & & \multicolumn{4}{c}{\textbf{Token-level}} & \multicolumn{4}{c}{\textbf{Concept-level}} & \textbf{Semantic-level} \\
\cmidrule(lr){3-6} \cmidrule(lr){7-10} \cmidrule(l){11-11}
\textbf{Dataset} & \textbf{Method} & \textbf{BLEU-1} & \textbf{BLEU-2} & \textbf{BLEU-3} & \textbf{BLEU-4} & \textbf{ROUGE-1} & \textbf{ROUGE-2} & \textbf{ROUGE-L} & \textbf{METEOR} & \textbf{RATE} \\ \midrule
\multirow{13}{*}{MIMIC-CXR} 
& Lingshu-7B$^{*}$ & 0.202 & 0.100 & 0.048 & 0.026 & 0.303 & 0.098 & 0.289 & 0.183 & 0.532 \\
& MedGemma-27B$^{*}$ & 0.228 & 0.113 & 0.052 & 0.019 & 0.247 & 0.061 & 0.233 & 0.215 & 0.520 \\
& HuatuoGPT-V-7B$^{*}$ & 0.226 & 0.100 & 0.035 & 0.010 & 0.230 & 0.045 & 0.214 & 0.196 & 0.489 \\
& HuatuoGPT-V-34B$^{*}$ & 0.225 & 0.096 & 0.035 & 0.010 & 0.231 & 0.045 & 0.216 & 0.199 & 0.490 \\
& Qwen2.5-VL-7B$^{\ddag}$ & 0.214 & 0.094 & 0.034 & 0.008 & 0.242 & 0.050 & 0.226 & 0.190 & 0.472 \\
& BiMediX2-8B$^{*}$ & 0.159 & 0.050 & 0.005 & 0.001 & 0.186 & 0.025 & 0.173 & 0.118 & 0.446 \\
& LLaVA-Med-7B$^{*}$ & 0.042 & 0.012 & - & - & 0.150 & 0.023 & 0.138 & 0.066 & 0.425 \\
& HealthGPT-14B$^{*}$ & 0.205 & 0.084 & 0.018 & 0.002 & 0.205 & 0.039 & 0.193 & 0.172 & 0.473 \\
& InternVL3-8B$^{\ddag}$ & 0.189 & 0.085 & 0.025 & 0.003 & 0.234 & 0.052 & 0.222 & 0.185 & 0.493 \\
& InternVL3-14B$^{\ddag}$ & 0.223 & 0.101 & 0.043 & 0.010 & 0.226 & 0.048 & 0.210 & 0.219 & 0.487 \\
& QoQMed-VL-7B$^{\S}$ & 0.122 & 0.056 & 0.022 & 0.006 & 0.185 & 0.037 & 0.174 & 0.188 & 0.495 \\
& MedVLM-R1-2B$^{\S}$ & 0.184 & 0.067 & 0.013 & 0.002 & 0.192 & 0.027 & 0.178 & 0.145 & 0.417 \\
\rowcolor[HTML]{F5F5F5} & HiMed-3B$^{\S}$ & 0.292 & 0.156 & 0.088 & 0.052 & 0.289 & 0.079 & 0.271 & 0.220 & 0.544 \\
\cmidrule(l){2-11}
\multirow{13}{*}{CheXpert} 
& Lingshu-7B$^{*}$ & 0.179 & 0.071 & 0.027 & 0.011 & 0.194 & 0.033 & 0.181 & 0.163 & 0.439 \\
& MedGemma-27B$^{*}$ & 0.117 & 0.047 & 0.018 & 0.006 & 0.165 & 0.033 & 0.153 & 0.183 & 0.471 \\
& HuatuoGPT-V-7B$^{*}$ & 0.130 & 0.047 & 0.016 & 0.003 & 0.157 & 0.026 & 0.145 & 0.180 & 0.435 \\
& HuatuoGPT-V-34B$^{*}$ & 0.093 & 0.031 & 0.010 & 0.002 & 0.151 & 0.024 & 0.141 & 0.151 & 0.431 \\
& Qwen2.5-VL-7B$^{\ddag}$ & 0.046 & 0.019 & 0.007 & 0.002 & 0.103 & 0.017 & 0.097 & 0.135 & 0.427 \\
& BiMediX2-8B$^{*}$ & 0.094 & 0.022 & 0.002 & - & 0.099 & 0.011 & 0.092 & 0.120 & 0.353 \\
& LLaVA-Med-7B$^{*}$ & 0.133 & 0.040 & 0.003 & - & 0.136 & 0.019 & 0.122 & 0.138 & 0.404 \\
& HealthGPT-14B$^{*}$ & 0.119 & 0.041 & 0.012 & 0.003 & 0.158 & 0.026 & 0.146 & 0.165 & 0.452 \\
& InternVL3-8B$^{\ddag}$ & 0.074 & 0.029 & 0.008 & 0.002 & 0.116 & 0.021 & 0.111 & 0.175 & 0.441 \\
& InternVL3-14B$^{\ddag}$ & 0.080 & 0.028 & 0.008 & 0.002 & 0.120 & 0.019 & 0.114 & 0.173 & 0.432 \\
& QoQMed-VL-7B$^{\S}$ & 0.033 & 0.015 & 0.005 & 0.001 & 0.090 & 0.014 & 0.086 & 0.115 & 0.454 \\
& MedVLM-R1-2B$^{\S}$ & 0.039 & 0.012 & 0.002 & 0.001 & 0.116 & 0.014 & 0.110 & 0.095 & 0.399 \\
\rowcolor[HTML]{F5F5F5} & HiMed-3B$^{\S}$ & 0.182 & 0.077 & 0.032 & 0.017 & 0.163 & 0.048 & 0.157 & 0.219 & 0.487 \\
\cmidrule(l){2-11}
\multirow{13}{*}{IU-Xray} 
& Lingshu-7B$^{*}$ & 0.362 & 0.203 & 0.130 & 0.085 & 0.386 & 0.138 & 0.359 & 0.312 & 0.590 \\
& MedGemma-27B$^{*}$ & 0.268 & 0.134 & 0.074 & 0.038 & 0.300 & 0.081 & 0.282 & 0.282 & 0.609 \\
& HuatuoGPT-V-7B$^{*}$ & 0.134 & 0.061 & 0.025 & 0.006 & 0.215 & 0.040 & 0.207 & 0.266 & 0.544 \\
& HuatuoGPT-V-34B$^{*}$ & 0.131 & 0.063 & 0.031 & 0.014 & 0.232 & 0.051 & 0.221 & 0.266 & 0.582 \\
& Qwen2.5-VL-7B$^{\ddag}$ & 0.057 & 0.027 & 0.011 & 0.004 & 0.158 & 0.031 & 0.154 & 0.181 & 0.569 \\
& BiMediX2-8B$^{*}$ & 0.094 & 0.018 & 0.002 & 0.001 & 0.100 & 0.009 & 0.096 & 0.134 & 0.420 \\
& LLaVA-Med-7B$^{*}$ & 0.116 & 0.036 & 0.003 & - & 0.140 & 0.018 & 0.131 & 0.142 & 0.432 \\
& HealthGPT-14B$^{*}$ & 0.126 & 0.053 & 0.022 & 0.010 & 0.207 & 0.040 & 0.193 & 0.227 & 0.541 \\
& InternVL3-8B$^{\ddag}$ & 0.082 & 0.037 & 0.018 & 0.007 & 0.149 & 0.034 & 0.142 & 0.233 & 0.561 \\
& InternVL3-14B$^{\ddag}$ & 0.100 & 0.042 & 0.017 & 0.004 & 0.166 & 0.032 & 0.158 & 0.242 & 0.556 \\
& QoQMed-VL-7B$^{\S}$ & 0.034 & 0.018 & 0.009 & 0.003 & 0.137 & 0.029 & 0.132 & 0.133 & 0.572 \\
& MedVLM-R1-2B$^{\S}$ & 0.184 & 0.067 & 0.013 & 0.001 & 0.153 & 0.021 & 0.149 & 0.145 & 0.531 \\
\rowcolor[HTML]{F5F5F5} & HiMed-3B$^{\S}$ & 0.407 & 0.259 & 0.171 & 0.115 & 0.320 & 0.089 & 0.319 & 0.411 & 0.611 \\ \bottomrule
\end{tabular}
}
\caption{The performance of different models on the MIMIC-CXR, CheXpert and IU-Xray datasets. Model types are separated into: $^{\ddag}$General Model, $^{*}$Medical Model, $^{\S}$RL-Based Model. Higher scores are better for all metrics.}
\label{tab:results} 

\end{table*}

\subsection{Main Results}
\label{sec:main_results}




We compare our proposed method with SOTA approaches, which are categorized into general-purpose, medical-domain, RL-based models, and MRG-specific methods. As shown in~\textbf{\textit{Table~\ref{tab:results}}}, our proposed HiMed-RL achieves the best performance across the majority of metrics. Notably, our model achieves the highest scores on the semantic-level RATE, indicating that it not only generates linguistically fluent and contextually relevant reports but, more importantly, accurately preserves the essential medical findings and impressions. Furthermore, our HiMed-RL model secures this SOTA performance with a relatively compact size of 3B parameters, surpassing various models with significantly larger parameter counts and thus highlighting the efficiency of our training strategy. 
Moreover, we compare our method with MRG-specific baselines. (Details analysis and results in \textbf{\textit{Appendix A.6}}) The results confirm their tendency to overfit on token-level details, which leads to a significant degradation in overall semantic quality.

\subsection{Ablation Study}
\begin{figure}[htbp]
    \centering
    \includegraphics[width=\linewidth]{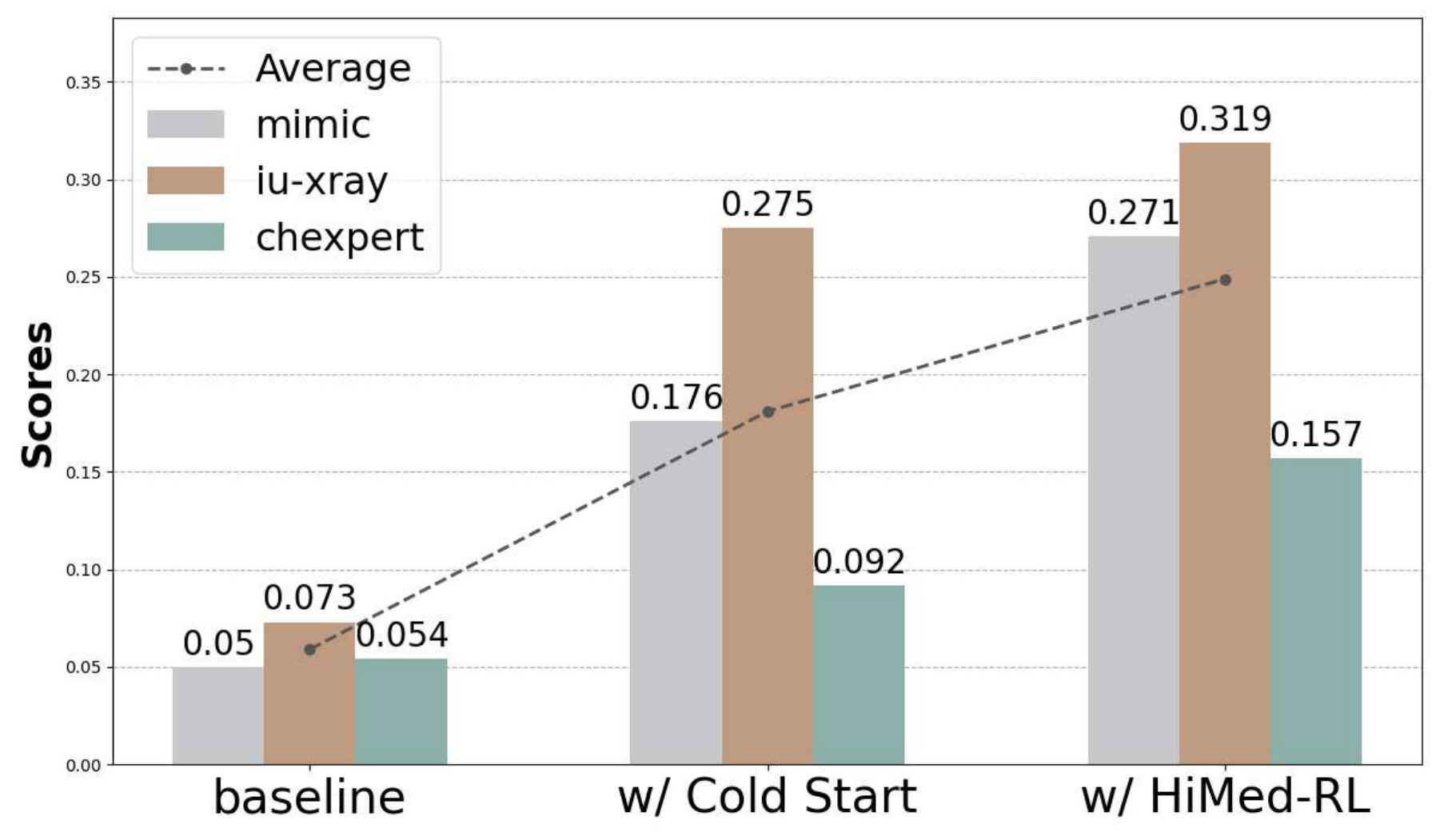}
    \caption{
    Ablation study results showcasing performance improvements in HiMed-3B across initial cold start and RL fine-tuning stages.
     }
    \label{fig:sft_rl_compare}
\end{figure}

Ablation studies are illustrated in \textbf{\textit{Figure~\ref{fig:sft_rl_compare}}}, we first evaluated the performance of the different training stages. We observed significant performance gains after both the initial cold-start SFT and the subsequent \mbox{HiMed-RL} fine-tuning. Notably, performance is enhanced during the RL phase, which demonstrates the powerful advantage of our training strategy in enhancing the quality of reports.
Next, our ablation study on the reward components in \textbf{\textit{Table~\ref{tab:ablation}}} reveals two key findings. First, each reward level individually provides an effective optimization signal, targeting distinct granularities: fluency (token), factual consistency (concept), and diagnostic logic (semantic). Second, combining these components demonstrates a strong synergistic effect, yielding the best performance across three datasets.

\begin{figure*}[h]
    \centering
    \includegraphics[width=0.9\linewidth]{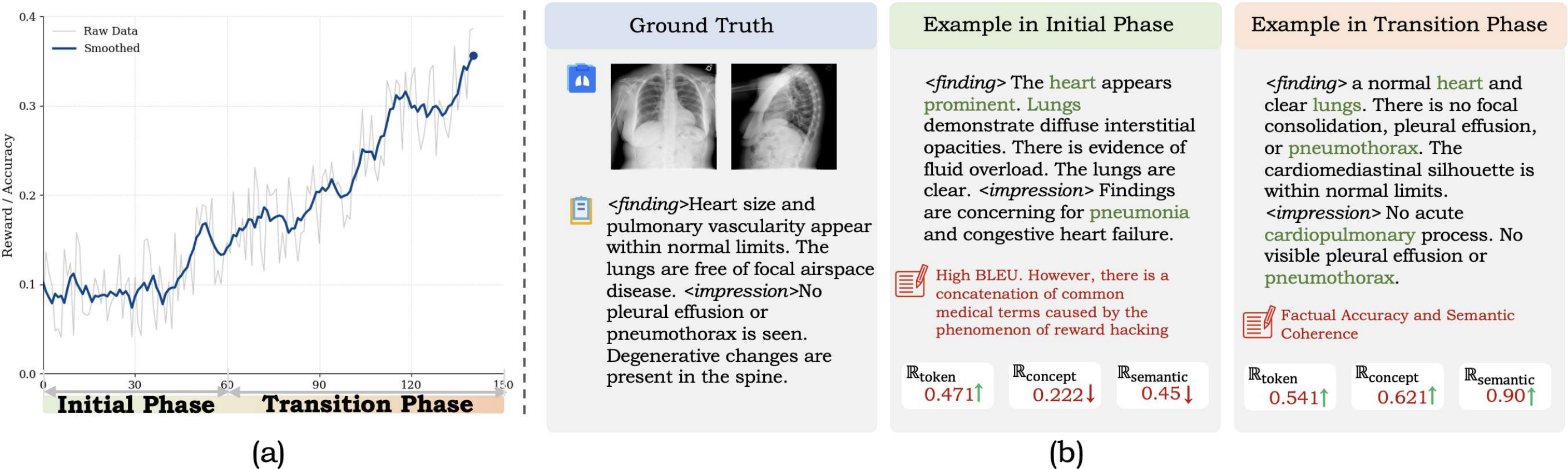}
    \caption{(a) Accuracy reward training curve. (b) Comparison of generated reports between the initial and transition phases.}
    \label{fig:case_study}
\end{figure*}


\subsection{Discussion}
\noindent \textbf{Human-inspired Dynamic Reward Adjustment Analysis.} We analyze the impact of this module's hyperparameters on HiMed-3B’s performance, as shown in \textbf{\textit{Table~\ref{tab:reward_adjustment_compact}}}. The fixed-weight configuration yields ROUGE-L of 0.264, METEOR of 0.211, and RATE of 0.525. In contrast, the optimal configuration ($T=10k, \alpha_{\text{min}}=0.1$) achieves ROUGE-L of 0.271, METEOR of 0.220, and RATE of 0.544, with improvements of 2.7\%, 4.3\%, and 3.6\%, respectively. These findings validate that a balanced transition from factual accuracy to semantic coherence, as facilitated by $T=10k$ and $\alpha_{\text{min}}=0.1$, best mimics clinical reasoning, enhancing report quality and mitigating hallucinations.

\begin{table}[htbp]
\centering
\small
\begin{tabular}{l S[table-format=1.3] S[table-format=1.3] S[table-format=1.3]}
\toprule
\textbf{Method} & {BLEU-1} & {METEOR} & {RATE} \\
\midrule
Qwen2.5VL-7B    & 0.024 & 0.067 & 0.348 \\
InternVL3-8B    & 0.010 & 0.004 & 0.027 \\
Lingshu         & 0.074 & 0.149 & 0.331 \\
\midrule
\textbf{HiMed-3B (Ours)} & \multicolumn{1}{c}{\bfseries 0.121\textbf{\textit{\textsubscript{+63.5\%}}}} & \multicolumn{1}{c}{\bfseries 0.178\textbf{\textit{\textsubscript{+19.5\%}}}} & \multicolumn{1}{c}{\bfseries 0.371\textbf{\textit{{\textsubscript{+12.1\%}}}}} \\
\bottomrule
\end{tabular}
\caption{
    Out-of-distribution performance on the Pad-Chest dataset.
    We compare HiMed-3B, against several strong baselines.
    Evaluation is conducted using: BLEU-1 ($\mathbb{R}_\text{token}$), METEOR ($\mathbb{R}_\text{concept}$), and RATE ($\mathbb{R}_\text{semantic}$).
}
\label{tab:padchest-results-revised}
\end{table}


\noindent \textbf{Generalization Study.} The generalization test is a core experiment in deep learning. For the bilingual Padchest-GR dataset, we specifically utilized the English subset to ensure linguistic consistency with our training data. As shown in \textbf{\textit{Table~\ref{tab:padchest-results-revised}}}, \mbox{HiMed-RL} achieves significant improvements against three strong baselines across all metrics, affirming its impressive generalization capabilities on unseen data. Results indicate that our strategy enables MLLMs to master intrinsic MRG patterns rather than rote memorization, ensuring practical value for real-world clinical applications.

\noindent \textbf{LLM Verifier Analysis.}
The effects of different LLM verifiers are presented in \textbf{\textit{Table~\ref{tab:judge_impact_simplified}}}. 
First, compared to the baseline without a verifier, introducing an LLM to provide semantic-level reward signals significantly enhances the quality of the final generated reports. 
Second, across all three datasets, \mbox{Qwen3-a3b (30B)} comprehensively outperforms the smaller models. 
This indicates that a more capable and larger-scale LLM, when acting as a ``semantic referee,'' provides more precise and effective reward signals, thereby improving report generation performance.

\subsection{Case Study}

\textbf{\textit{Figure~\ref{fig:case_study} (b)}} presents a case comparison that highlights the efficacy of our proposed adjustment strategy. 
In the initial training phase, the generated report achieves a high token-level reward ($\mathbb{R}_\text{token}$), yet it contains severe contradictions and factual errors. 
For instance, it simultaneously claims the presence of ``diffuse interstitial opacities'' and that the ``lungs are clear.'' 
Furthermore, the impression of ``pneumonia and congestive heart failure'' is entirely inconsistent with the ground truth.
We attribute this phenomenon to reward hacking, wherein the model naively stacks professional medical terminology to maximize superficial rewards, while disregarding semantic coherence and factual alignment. 
\begin{table}[htbp]
    \centering   
    \small
    \setlength{\tabcolsep}{4pt}
    \begin{tabularx}{\columnwidth}{>{\centering\arraybackslash}X c |ccc}

        \toprule
        \textbf{LLM verifier} & \textbf{Params} & 
         MIMIC-CXR & CheXpert & IU-Xray \\
        \midrule
        \textit{w/o $\mathbb{R}_\text{semantic}$} & -- & 0.258 & 0.138 & 0.301 \\
        \midrule
        Llama3.2 & 8B & 0.267 & 0.134 & 0.311 \\
        Qwen2.5-I & 7B & 0.266 & 0.135 & 0.313 \\
        InternVL2 & 7B & 0.265 & 0.133 & 0.310 \\  \midrule
        Qwen3-a3b & 30B & \textbf{0.271} & \textbf{0.157} & \textbf{0.319} \\
        \bottomrule
    \end{tabularx}
    \caption{Analysis on the LLM verifier for the semantic-level reward ($\mathbb{R}_\text{semantic}$). The best judge verifier is highlighted. We use Rouge-L to compare the performance.}
\label{tab:judge_impact_simplified}
\end{table}

Turning to the transition phase, after adjusting the weights of the semantic-level rewards, the generated report becomes coherent and highly consistent with the ground truth, achieving high scores across all three reward levels. 
This comparison reveals the essence of our hierarchical reward learning paradigm: by optimizing these reward signals, we guide the model beyond superficial mimicry towards a genuine understanding of the medical concepts within the image, thereby producing logically sound and factually accurate reports.

The reward curve in \textbf{\textit{Figure~\ref{fig:case_study} (a)}} corroborates this qualitative analysis. The model's accuracy reward consistently improves throughout training. After a period of slow and unstable improvement in the initial phase, the performance accelerates significantly during the transition phase. 

\section{Conclusion}
We proposed \mbox{HiMed-RL}, a hierarchical reward learning framework, which leverages reward signals from multiple linguistic granularities to guide the generation of high-quality reports, and its effectiveness is demonstrated through comprehensive experiments.
The insight of our work is that by coordinating complementary reward signals with a dynamic learning strategy, we can impel an MLLMs to transition from superficial mimicry to genuine understanding. 

\section{Acknowledgments}
This work was supported by the National Key R\&D Program of China (Grant No. 2024YFC3308304), the ``Pioneer'' and ``Leading Goose'' R\&D Program of Zhejiang (Grant No. 2025C01128), the National Natural Science Foundation of China (Grant No. 62476241), and the Natural Science Foundation of Zhejiang Province, China (Grant No. LZ23F020008) and the ZJU-Angelalign R\&D Center for Intelligence Healthcare.

\end{document}